\documentclass[aps,prl,twocolumn,longbibliography,reprint]{revtex4-2}

\usepackage[colorlinks,hyperindex,plainpages=false,allcolors={blue}]{hyperref}
\usepackage{longtable}
\usepackage{morefloats}
\usepackage{color}
\usepackage{graphicx}
\usepackage[utf8]{inputenc}
\usepackage{graphicx}
\usepackage{multirow}
\usepackage{hhline}
\usepackage{natbib}

\usepackage{amsmath,amsfonts,amsthm,amssymb}
\usepackage{appendix}
\usepackage{makeidx}
\usepackage{url}
\usepackage{verbatim}
\usepackage[rightcaption]{sidecap}
\usepackage{array}
\usepackage{booktabs}
\usepackage{multirow}
\usepackage{tabularx}
\usepackage{mathrsfs}
\usepackage{soul,xcolor}
\usepackage{times}
\usepackage{txfonts}
\usepackage{bm} 
\usepackage{soul}
\usepackage{orcidlink}

\def  \bOm    {\mbox{\boldmath$\Omega$}}

\begin{document}

\title{Exchange and spin-orbit proximity driven topological and transport phenomena in twisted {graphene/CrI$_3$} heterostructures}

\author{M. Jafari$^{1 \ast}$,  M. Gmitra$^{2,3}$, A. Dyrda\l$^{1}$}
\email{adyrdal@amu.edu.pl} \email{mirali.jafari@amu.edu.pl}
\affiliation{$^{1}$Faculty of Physics and Astronomy, ISQI,
Adam Mickiewicz University in Poznań,\\ ul. Uniwersytetu Poznanskiego 2, 61-614 Pozna\'n, Poland\\
$^{2}$Institute of Physics, Pavol Jozef Safarik University in Kosice, 04001 Kosice, Slovakia\\
$^{3}$Institute of Experimental Physics, Slovak Academy of Sciences, 04001 Kosice, Slovakia}
\vspace{10pt}

\date{\today}
\begin{abstract}
We present results of comprehensive first-principles and kp-method studies of electronic, magnetic, and topological properties of graphene on a monolayer of CrI$_3$. First, we identify a twist angle between the graphene and CrI$_3$, that positions the graphene Dirac cones within the bandgap of CrI$_3$. Then, we derive the low-energy effective Hamiltonian describing electronic properties of graphene Dirac cones.  Subsequently, we  examine anomalous and valley Hall conductivity and discuss possible topological phase transition from a quantum anomalous Hall insulator to a trivial insulating state, concomitant a change in the magnetic ground state of CrI$_3$. These findings highlight the potential of strain engineering in two-dimensional van der Waals heterostructures for controlling topological and magnetic phases.
\end{abstract}
\maketitle

\section{Introduction}
Since the discovery of graphene~\cite{novoselov2004electric}, there has been remarkably rapid progress in synthesizing two-dimensional (2D) materials, assembling them into van der Waals (vdW) heterostructures, and in their functionalization~\cite{geim2013van,Gong_2DMagnetsRev_Science2019,Kim_Song_ACSMatAu2022,Castellanos-Gomez_NatRevMetPrim2022,Bian_NSR2022,Qi_rev_2023}. 
Over the past two decades, advances in fabrication  of vdW materials  coincided with the discovery of novel phenomena and topological phases driven by the choice of 2D crystals and intentional  twisting of  layers.~\cite{Fert_Panagopoulos_Nature2016,Tokura_NatPhys2017,Gong_2DMagnetsRev_Science2019,Bao_NatPhysRev2022,Kurebayashi_NatureRev2022,  Bernevig_NatRevPhys2022,MagneticGenomevdW_ACSNano2022,Zhang_Small_2022,Fert_RevModPhys2024,ren20252dmaterialsroadmap,Sun_ChemRev2024,Zhang_2DMagnets_rev_npjSpintronics_2024}. In addition, vdW heterostructures offer exceptional tune-ability by electric and magnetic fields, mechanical forces, and temperature, making them a unique platform for both fundamental research—spanning solid-state physics, magnetism, quantum transport, and optics, as well as for emerging applications in spintronics, data storage, quantum technologies, sensors, and beyond~\cite{Vedmedenko_2020,Kurebayashi_NatureRev2022,Valenzuela_Roche_MRAMs_nature2022,optospin-NatureElectronics2024,Becher_2023}.

Within the rapidly growing family of van der Waals materials, graphene-based vdW heterostructures remain the most intensively studied. When combined with semiconducting transition metal dichalcogenides (TMDCs)~\cite{{Gmitra_Fabian_PRB2015,Gmitra_Fabian_PRB2016,Cumings_Fabian_PRL2017,Ghiasi_NanoLett2017,Garcia_Roche_RSC_ChemSocRev2018,Wakamura_PhysRev2019,Ghiasi_NanoLett2019}}, graphene exhibits strong spin–orbit proximity effects, leading to spin-to-charge conversion phenomena such as the spin Hall effect and current-induced spin polarization. In turn, graphene on ferromagnetic layers exhibits proximity-induced exchange effects, showing the anomalous Hall effect up to room temperature and the quantum anomalous Hall state when the Fermi level is within the band gap~\cite{Kawakami_ACSNano2012,Yang_Roche_Chshiev_PRL2013,GrEuO_PhysRevB.95.075418,Hallal_2DMaterials2017,Averyanov_ACSApplMatInt2018,doi:10.1126/sciadv.aax9989,Zollner_PRL2022,Pandey_PRB2023}.

Graphene also seems to be an ideal platform for developing the twistronics~\cite{Liu_PhysRev2019,Carr_NatRevMat2020,he2021moire,Liu_Wang_AdvSci_moire2022,Aggarwal_2023}. Twist-angle-dependent spin–orbit proximity effects in graphene on TMDCs have been explored both theoretically and experimentally.~\cite{Yang2019,DavidBurkard2019,NaimerFabian2021,BurkardKormanyos_2022,Casanova_PRB22,Pezo_2022,Zollner_PRB2023,PhysRevApplied.19.014053,Wojciechowska_arxiv25}. Apart from this, twist-angle-tunable magnetism and magnetic proximity effects are currently attracting considerable attention.
In multilayer graphene systems, that can serve as orbital Chern insulators, the electric-field control of magnetic states has been experimentally demonstrated.~\cite{Polshyn_Watanabe_Taniguchi_Nature2020}. Moreover,   a twist-angle-induced magnetism  in twisted bilayer WSe$_2$ has been reported, recently~\cite{Adam_arxiv2025}.
Twisting has also been shown to induce nonrelativistic spin splitting in antiferromagnetically coupled bilayers and to generate large transverse spin currents under external electric fields without spin-orbit coupling~\cite{He_PRL2023}. Interestingly, altermagnetism has recently been observed in twisted magnetic van der Waals bilayers, opening a new avenue in altermagnetic spintronics, with  twist-angle-controlled large spin currents~\cite{Liu_PRL2024}. 
The twist angle also provides a new degree of freedom to tune the proximity exchange coupling in graphene deposited on magnetic van der Waals crystals, controlling both its strength and its character from ferromagnetic to antiferromagnetic~\cite{Yan_PhysicaE2021,Zollner_PRL2022}. 

\begin{figure}[t]
 \centering
\includegraphics[width=0.95\columnwidth]{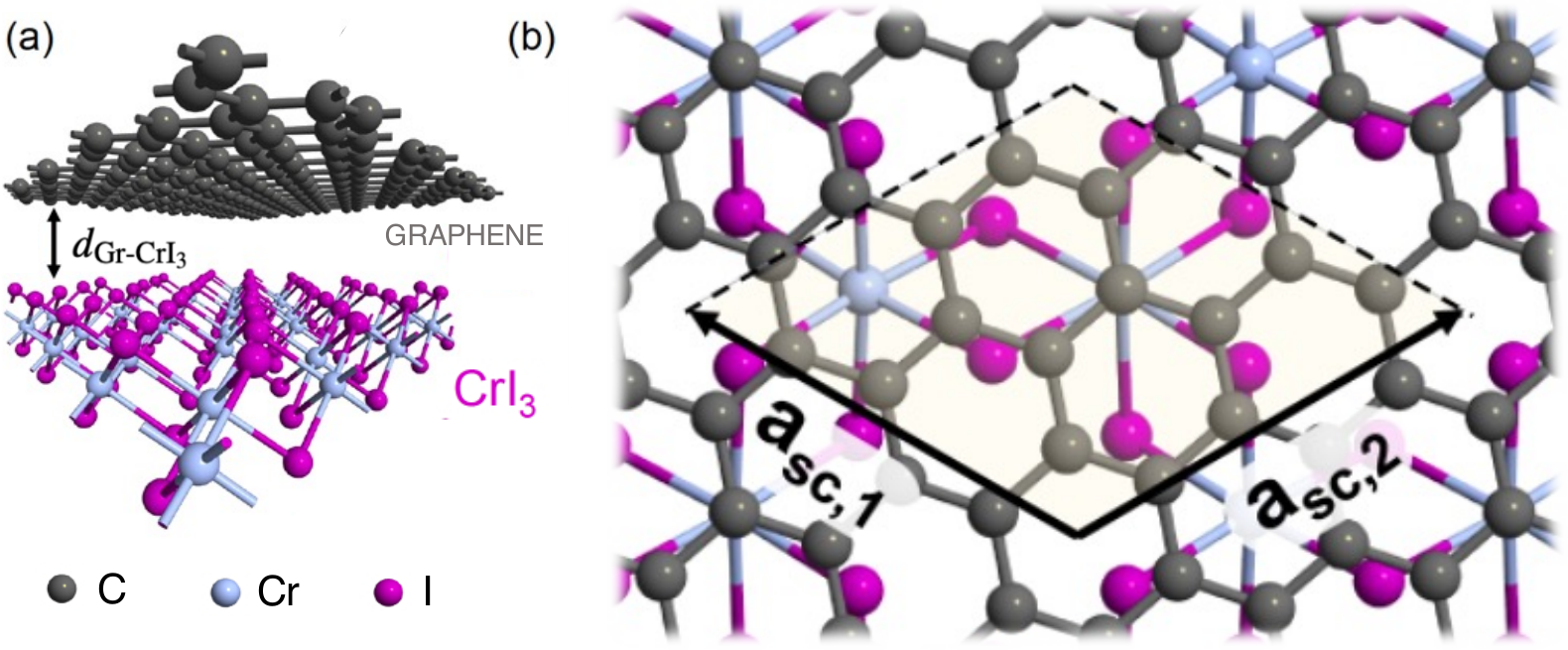}
 \caption{Structural model of the t-Gr/CrI$_3$ heterostructure: (a) side view and (b) top view, with the corresponding supercell outlined.}
\label{fig:Fig1}
\end{figure}

Transition metal trihalides are among the most intensively studied magnetic van der Waals crystals~\cite{AtlasTMTrihalides,Zhang_2DMagnets_rev_npjSpintronics_2024}. Cr-based trihalides, CrX$_3$ ($\mathrm{X = I, Cl, Br}$), are insulating, exfoliable 2D magnets, whose magnetic interactions depend strongly on stacking. Their magnetic ground state and anisotropy energy -- both in monolayers and multilayers -- can be tuned by magnetic and electric fields or by strain and pressure \cite{Webster_PRB2018,Huang_NatNano2018,Leon_2DMat2020,Ebrahimian_SciRep2023}.

The electronic and magnetic properties of single- and bilayer graphene/CrX$_3$ heterostructures have recently been investigated both theoretically and experimentally~\cite{Tseng_Watanabe_Yankowitz_NanoLett2022,Tenasini_Watanabe_Taniguchi_NanoLett2022,Farooq}. It has been shown that electronic states of graphene overlap and strongly hybridize with CrI$_3$ orbitals~\cite{Farooq}. However, the application of gate voltage can drastically change the electronic band structure, shifting graphene electronic states to the bandgap of CrI$_3$. The hydrostatic pressure applied to graphene/CrI$_3$ can have a similar effect~\cite{Zhang_PRB2018}. Importantly, pressure and gate voltage allow controlling the bandgap width in electronic states of graphene~\cite{https://doi.org/10.1002/apxr.202300026}. Moreover, the above-mentioned first-principles results indicate valley-contrasting physics in graphene deposited on CrI$_3$. 

The ferromagnetic proximity exchange effect in graphene on CrI$_3$ makes graphene a Chern insulator with quantized anomalous Hall conductivity, when the chemical potential is within the energy gap. The ability to control its electronic and topological properties by external means, such as gate voltage and pressure -- together with the system’s intrinsic valley-contrasting physics -- makes graphene/CrI$_3$ a compelling platform for exploring new topologically nontrivial phases and phenomena.

In this paper, we propose an alternative way to shift the electronic states of graphene into the bandgap of CrI$_3$ in graphene/CrI$_3$ (Gr/CrI$_3$) heterostructures. We have found that the twist of the graphene layer with respect to the CrI$_3$ layer by the angle $\Theta = 10.89^{\circ}$ places the graphene Dirac cones close to the Fermi level. The low-energy electronic states of the twisted graphene on CrI$_3$ (t-Gr/CrI$_3$) reveal a band inversion -- a characteristic signature of systems with nontrivial topological order. Indeed, we found that the system is a Chern insulator with quantized anomalous Hall conductivity when all valence bands are occupied. Furthermore, we analyse in this paper the behavior of selected transport characteristics, including the anomalous and valley Hall effects and the anomalous and valley Nernst effects.
We also consider the effect of lateral biaxial strain on CrI$_3$, which can induce a magnetic phase transition to an antiferromagnetic ground state, accompanied by a topological phase transition to a trivial insulating phase.
Our results show the possibility of controlling the band structure in graphene-based heterostructures {\it via} the twist angle. In turn, the interplay of twisting and mechanical strain in twisted graphene-based heterostructures provides a new route towards tunable topological spintronics.

\section{Structure and first-principles modelling}

Structure model of the considered t-Gr/CrI$_3$ heterostructure is shown in Fig.~\ref{fig:Fig1}, where
Fig.~\ref{fig:Fig1}(a) shows the side view, while Fig.~\ref{fig:Fig1}(b) presents the top view of the heterostructure, with the corresponding supercell outlined. The lattice parameter of pristine graphene is equal to 2.46~\AA, while that of CrI$_3$  is 6.92~\AA . To perform first-principles calculations using the Bloch theorem, 
we define commensurate supercells for the heterostructures.
Accordingly, using the coincidence lattice method, we consider below the system geometry and determine parameters of the supercells suitable for density functional theory (DFT) calculations. Then,  we briefly outline technical details of the DFT calculations, and -- at the end of this section -- we present and discuss the electronic band structure and magnetic properties of the t-Gr/CrI$_3$ heterostructure.

\begin{figure*}
 \centering
\includegraphics[width=0.99\textwidth]{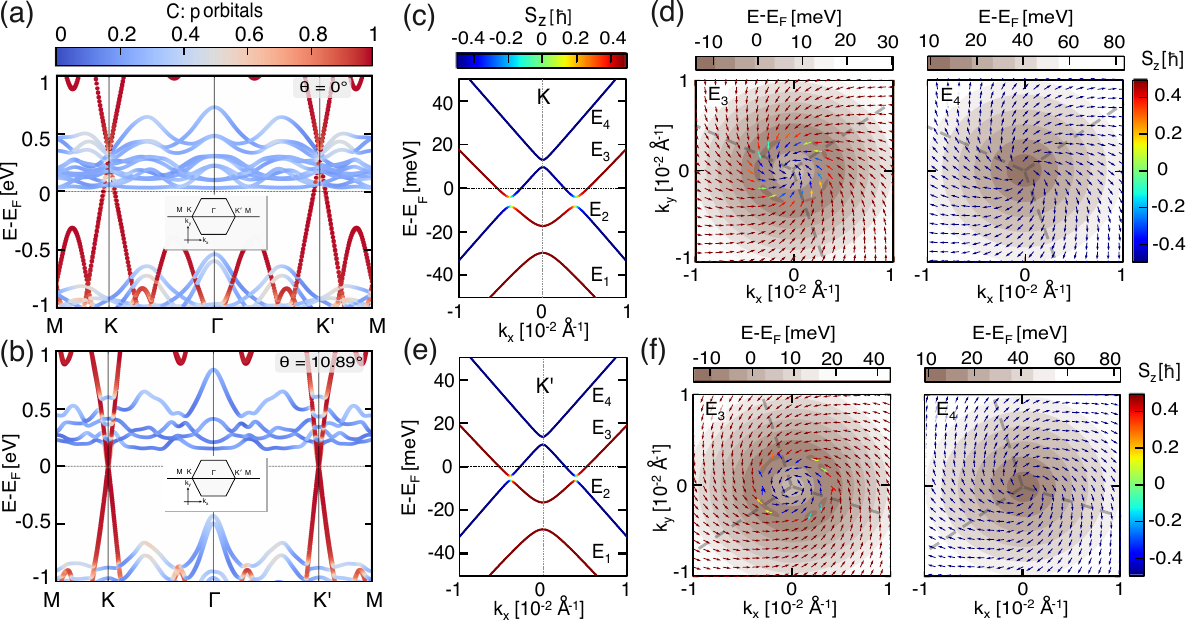}
 \caption{Band structure  of Gr/CrI$_3$ from DFT calculations. (a),(b) The band structure along the {\rm M}–{\rm K}–{\rm $\Gamma$}–{\rm K}'–{\rm M} path for $\Theta = 0^{\circ}$ (a), and $\Theta = 10.89^{\circ}$ (b). The colour scale presents the weighted contribution of Carbon $p$ orbitals to each band.
 (c),(e) Band structures of t-Gr/CrI$_3$ zoomed on graphene electronic states around the K and K' points of the Brillouin zone. The colour scale represents the corresponding spin $S_z$ expectation value. (d),(f) In-plane spin expectation values in the wavevector space, plotted in the vicinity of the K and K' points (the point $(k_x, k_{y}) = (0, 0)$ defines K/K' point, respectively).}
\label{fig:Fig2}
\end{figure*}

\subsection{Supercell Geometry}
We have performed first-principles calculations for both untwisted and twisted heterostructures consisting of graphene on CrI$_3$ monolayer. To construct the relevant supercell, we use the coincidence lattice method ~\cite{Wang_JPhysChem2015,Koda_JPhysChem2016,Carr_NatRev2020,Zollner_Kurpas_Gmitra_Fabian_NatRev2025}. Assuming that $\mathbf{a}_{1,2}$ are the vectors defining the primitive
unit cell of graphene, and $\mathbf{b}_{1,2}$ are the vectors defining the primitive
unit cell of CrI$_3$, one can define their supercells by the vectors $\mathbf{a}_{sc,1,2}^{(n,m)}$, and $\mathbf{b}_{sc,1,2}^{(n',m')}$, respectively:
\begin{eqnarray}
\mathbf{a}_{sc,1}^{(n,m)} = n\, \mathbf{a}_{1} + m\, \mathbf{a}_{2}, \,\,\,\,\,\,\,\,\,\,
\mathbf{a}_{sc,2}^{(n,m)} = \hat{R}_{\theta} \,\mathbf{a}_{sc,1}^{(n,m)}\,\,\,\,\,\, \\
\mathbf{b}_{sc,1}^{(n',m')}  = n'\, \mathbf{b}_{1} + m'\, \mathbf{b}_{2}, \,\,\mathbf{b}_{sc,2}^{(n',m')}  = \hat{R}_{\theta'} \,\mathbf{b}_{sc,1}^{(n',m')}
\end{eqnarray}
where $\{n, m, n', m'\} \in \mathbb{Z}$, while $\hat{R}_{\theta}$ and  $\hat{R}_{\theta'}$
are the rotation matrices by the angle $\theta :=\measuredangle(\mathbf{a}_{1}, \mathbf{a}_{2})$ and  $\theta' :=\measuredangle(\mathbf{b}_{1}, \mathbf{b}_{2})$ respectively. As both graphene and CrI$_{3}$ form the hexagonal lattices, with the lattice constant $a = |\mathbf{a}_{1}|= |\mathbf{a}_{2}|$ and $b = |\mathbf{b}_{1}|= |\mathbf{b}_{2}|$, respectively, one can choose $\theta = \theta' = 60\deg$.
Accordingly, the relative twist angle of the supercell with respect to the primitive unit cell for graphene is given by the formula 
\begin{equation}
\Theta = \theta_{0} - \arctan\left(\frac{\sqrt{3} m'}{2 n' + m'}\right) - \arctan\left(\frac{\sqrt{3} m}{2 n + m}\right),
\end{equation}
where $\theta_{0}$ is the initial orientation between lattices 
that in our case is 30$^\circ$.

In order to construct the commensurate supercells for the whole vdW heterostructure, as required for
periodic DFT calculations, one needs to construct the individual supercells in a way that $a_{sc}^{(n,m)}\approx b_{sc}^{(n',m')}$, and then to apply a mechanical strain to one of the supercells to fulfil the condition $a_{sc}^{(n,m)}= b_{sc}^{(n',m')}$.
We consider two supercell geometries: in the first case, the graphene monolayer is placed directly on top of a CrI${_3}$ monolayer without any twist; in the second case, the graphene monolayer is twisted by the angle of $\Theta = 10.89^{\circ}$ on the fixed monolayer of CrI${_3}$.
Table~\ref{tab:tab1} collects the parameters defining lattice vectors of the supercells, the number of atoms constituting the supercells, and the calculated biaxial strain applied to graphene. 
 \begin{table}[t]
  \caption{Structural and computational parameters for the studied systems.}
    \centering
    \renewcommand{\arraystretch}{1.2} 
    \setlength{\tabcolsep}{4pt}       
    \begin{tabular}{
        @{}
        >{\centering\arraybackslash}p{0.085\linewidth} 
        >{\centering\arraybackslash}p{0.095\linewidth} 
        >{\centering\arraybackslash}p{0.095\linewidth} 
        >{\centering\arraybackslash}p{0.065\linewidth} 
        >{\centering\arraybackslash}p{0.105\linewidth} 
        >{\centering\arraybackslash}p{0.065\linewidth} 
        >{\centering\arraybackslash}p{0.12\linewidth}  
        >{\centering\arraybackslash}p{0.12\linewidth}  
        >{\centering\arraybackslash}p{0.07\linewidth}  
        @{}
        }
    \hline
    $\Theta$ & $(n,m)$ & $(n',m')$ & $N$ & $\varepsilon_{\mathrm{Gr}}$ [\%] & $n_{k}$ & $a_{sc}$ [$\text{\AA}$] & $d$ [$\text{\AA}$] & GS \\
    \hline
    0$^{\circ}$      & (5,0) & (1,1) & 74 & $-2.60$ & 8  & 11.99 & 3.4901 & FM \\
    10.89$^{\circ}$  & (2,1) & (1,0) & 22 & $+6.27$ & 12 & 6.92  & 3.4829 & FM \\
    \hline
    \end{tabular}
    \label{tab:tab1}
\end{table}
\subsection{Computational details}
The first-principles calculations were performed within density-functional theory (DFT) \cite{hohenberg1964inhomogeneous}. The exchange–correlation energy was treated in the generalized-gradient approximation (GGA) using the Perdew–Burke–Ernzerhof (PBE) functional \cite{perdew1996generalized}. A plane-wave basis was employed together with scalar-relativistic projector-augmented-wave (PAW) pseudopotentials~\cite{kresse1999ultrasoft} as implemented in the \textsc{Quantum~Espresso} package~\cite{giannozzi2009quantum}. To account for spin–orbit coupling (SOC), we used fully relativistic PAW datasets. The kinetic-energy cutoff for wavefunctions was set to 70~Ry, whereas the charge-density energy cutoff was set to 500~Ry. Long-range interactions between graphene and the CrI$_3$ monolayer was described using the Grimme DFT-D2 dispersion correction~\cite{grimme2006semiempirical}. Electronic occupations followed a Marzari-Vanderbilt smearing with width $k_{\mathrm B}T=0.0005$~Ry \cite{marzari1999thermal}.
In turn, on-site electron correlations for the Cr $3d$ states were included via the DFT+$U$ (U = 3 eV) approach of Cococcioni~\cite{cococcioni2005linear}. To eliminate interactions between periodic images along the $c$ axis, a vacuum spacing of 25~\AA\,  has been added. Finally,  Brillouin-zone integrations have been performed employing a uniform $n_k \times n_k \times 1$ $k$-point mesh, with the $n_k$ values for each geometry listed in Table~\ref{tab:tab1}. 

\subsection{Electronic and magnetic properties of Gr/CrI$_3$ heterostructure}

Pristine CrI$_3$ is a ferromagnet with the easy axis oriented perpendicularly to the monolayer plane. The magnetic anisotropy energy (MAE) is defined as the difference in total energy between the out-of-plane and in-plane magnetization orientations: $\mathrm{MAE = E_{[100]} - E_{[001]}}$. For the pristine  CrI$_3$ monolayer we found  ${\mathrm{MAE_{CrI_3} = 0.869~meV}}$,  that is in good agreement with other literature data~\cite{Webster_PRB2018,Zhang_PRB2018}. However, deposition of graphene on CrI$_3$ leads to a remarkable decrease in MAE, and we found ${\mathrm{MAE_{Gr/CrI3}(\Theta=0^{\circ}) = 0.476meV}}$ for the untwisted Gr/CrI$_3$ structure, and   $\mathrm{MAE_{tGr/CrI_{3}}(\Theta=10.89^{\circ}) = 0.71546~meV}$ for the twisted structure.
This indicates that the electronic states of graphene on CrI$_3$ are influenced by both
proximity-induced exchange field and spin-orbit coupling, which lift the
degeneracy of the graphene $p$-states and deforms the Dirac cones in the K and K’ points.

Figures~\ref{fig:Fig2}(a,b) show  the band structure of Gr/CrI$_3$ obtained from  DFT calculations, plotted along the high-symmetry path {\rm M}–{\rm K}–{\rm $\Gamma$}–{\rm K}'–{\rm M} in the Brillouin zone, and  for the twist angles $\Theta = 0^{\circ}$ (a)  and $\Theta = 10.89^{\circ}$ (b), respectively. The presented data correspond to the tensile strain $\varepsilon = +6.27\%$ applied to graphene (and zero strain, $\varepsilon_{\mathrm{CrI_3}} = 0$, on CrI$_3$) to ensure commensurability of the DFT calculations. 
For the untwisted heterostructure, Fig.~\ref{fig:Fig2}(a), the graphene Dirac cones merge with the electronic band of CrI$_3$. In turn, for the twisted structure, the Dirac cones are shifted to the bandgap of CrI$_3$, see Fig.~\ref{fig:Fig2}(b). 
Figures~\ref{fig:Fig2}(c,e) present the low-energy electronic states of graphene in the vicinity of the K and K' Dirac points,  modified by the proximity effects.  
In graphene deposited on CrI$_3$, the time-reversal symmetry, spatial inversion symmetry, and sublattice (pseudospin) symmetry are broken. Consequently, one observes the interplay between exchange field, spin-orbit coupling, and staggered potential, which altogether lead to spin splitting and anticrossing of the bands and to a finite band gap. These proximity effects will be elaborated in more detail in the subsequent section. To complete the DFT results, in Figures~\ref{fig:Fig2}(d,f) we show the spin expectation values, corresponding to the states shown in  Figures~\ref{fig:Fig2}(c,d). The most important feature of the results shown in Figures~\ref{fig:Fig2} (c - f) is the inverted band structure — a hallmark of nontrivial topological properties. Moreover, one can also note the valley-contrasting behaviour, i.e.,  the bandgap at the K point is {4.15 meV} whereas at the K' point it is {2.15~meV}.  

\begin{figure*}[ht]
 \centering
\includegraphics[width=0.9\textwidth]{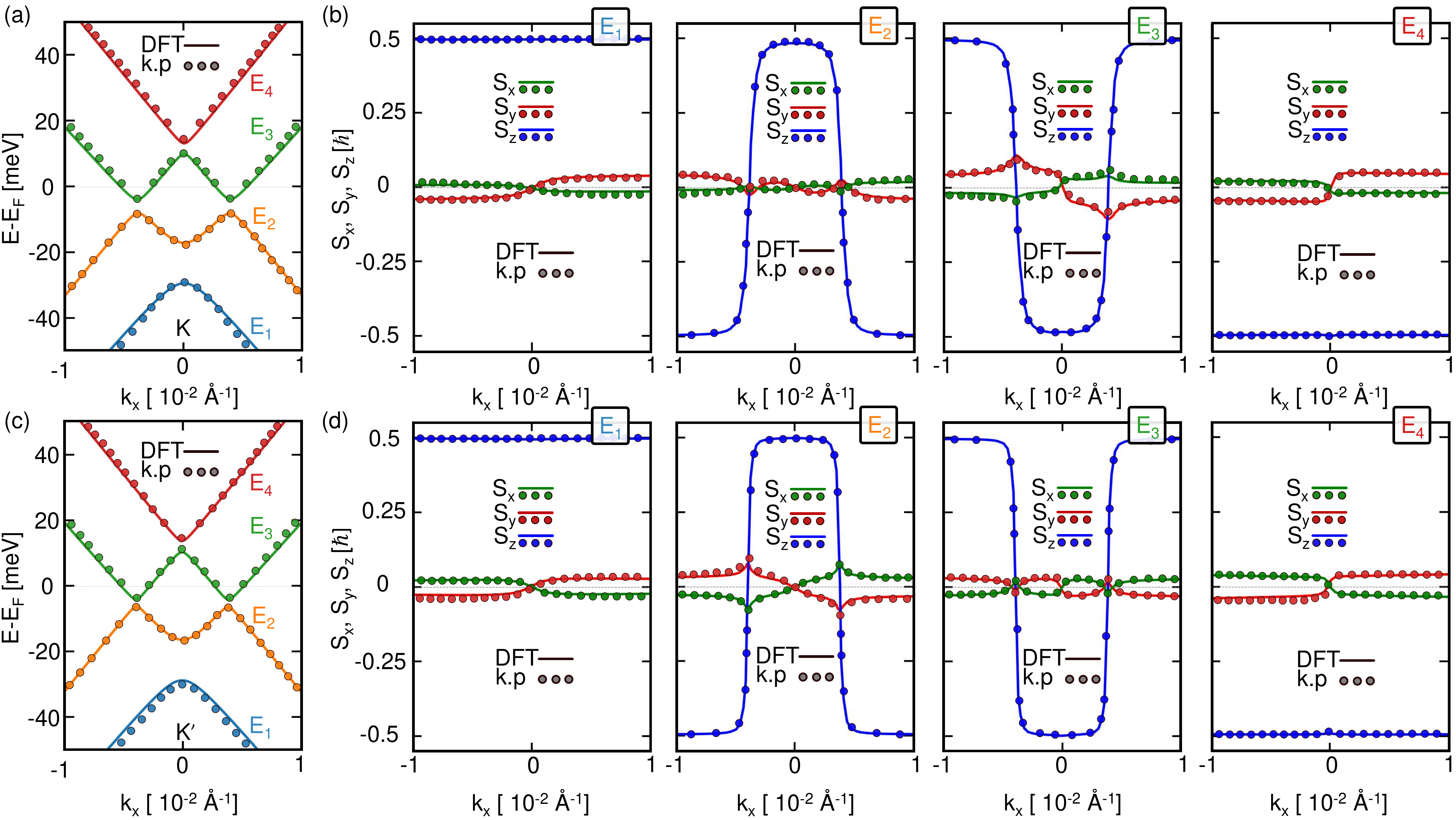}
 \caption{Electronic band structure of tGr/CrI$_3$ as well as the $S_x$, $S_y$, and $S_z$ components of spin expectation values in the vicinity of the K and K' points. Solid lines represent the DFT results, while the dots correspond to the results obtained from the $\mathbf{k}\cdot\mathbf{p}$ model Hamiltonian.}
\label{fig:Fig1sm}
\end{figure*}

\section{Low-energy effective model}
The electronic states of proximitized graphene can be described within an effective tight-binding Hamiltonian that incorporates the symmetries of local atomic orbitals~\cite{KochanFabian2017}. In the vicinity of the graphene K and K' points, this Hamiltonian can be expanded according to the standard $\mathbf{k}\cdot\mathbf{p}$ method. As a result, the physics of the low-energy electronic states of graphene can be described by an effective low-energy $\mathbf{k}\cdot\mathbf{p}$ Hamiltonian, whose explicit matrix form can be determined by taking into account the relevant symmetry arguments and also features of the DFT band structure. 
When expressed in terms of the Pauli matrices acting in the sublattice, $\hat{\sigma}_{\alpha}$, and spin, $\hat{s}_{\alpha}$ subspaces ($\alpha = \{0, x,y,z\}$, with $\alpha = 0$ corresponding to the identity $2\times 2$ matrix), this Hamiltonian takes the form~\cite{Kane_Mele_PRL2005,KochanFabian2017,Phong_Guinea_2018,Zollner_PRB2016,Hogl_Gmitra_PRL2020,Yang2019,DavidBurkard2019,BurkardKormanyos_2022}:
\begin{equation}
\label{eq:H_kp}
\hat{H}_{\tau} = \hat{H}_{0}^{\tau} + \hat{H}_{\Delta}+ \hat{H}_{I}^{\tau}  + \hat{H}_{R}^{\tau} + \hat{H}_{EX}^{\tau} + E_{0} \hat{\sigma}_{0}\otimes\hat{s}_{0},
\end{equation}
where $\tau = \pm1$ corresponds to the K/K' valley, respectively. The first term describes the low-energy states of pristine graphene~\cite{KochanFabian2017,Phong_Guinea_2018}:
\begin{equation}
\hat{H}_{0}^{\tau} = v \left(\tau \hat{\sigma}_{x} k_{x} + \hat{\sigma}_{y} k_{y} \right),
\end{equation}
with the parameter $v = a t\frac{\sqrt{3}}{2}$ defining the Fermi velocity, and determined by the lattice constant $a$ and nearest neighbour hopping parameter $t$. The second term is responsible for the orbital proximity-induced gap arising from the pseudospin symmetry breaking: carbon atoms on sublattices A and B experience, on average, different crystalline fields due to the presence of CrI$_3$. This term takes the form~\cite{KochanFabian2017,Phong_Guinea_2018}:
\begin{equation}
\hat{H}_{\Delta} = \Delta \hat{\sigma}_{z}\otimes\hat{s}_{0} ,
\end{equation}
where $\Delta$ is the so-called staggered potential (effective orbital hybridization energy) on sublattices A and B. 
The effect of sublattice-dependent spin-conserving next-nearest neighbor spin-orbit interaction is described by the term~\cite{Kane_Mele_PRL2005,KochanFabian2017,Phong_Guinea_2018}:
\begin{equation}
\hat{H}_{I}^{\tau} = \tau \left(\lambda_{I}^{A} \hat{\sigma}_{+}  + \lambda_{I}^{B} \hat{\sigma}_{-}\right) \otimes \hat{s}_{z}
\end{equation}
where $\hat{\sigma}_{\pm} = (\hat{\sigma}_{z} \pm \hat{\sigma}_{0})/2$, and $\lambda_{I}^{A/B}$ denotes the intrinsic spin-orbit parameter for
the sublattices A/B, respectively. 
Due to the spatial inversion symmetry breaking, the Rashba spin-orbit interaction appears in the system and tilts electrons' spin out of the plane. The Rashba Hamiltonian in twisted graphene-based structures takes more general form~\cite{Kane_Mele_PRL2005,KochanFabian2017,Phong_Guinea_2018,Yang2019,DavidBurkard2019,BurkardKormanyos_2022}: 
\begin{equation}
\hat{H}_{R}^{\tau} = - \lambda_{R} \exp^{- i \frac{\phi_{R}}{2} \hat{s}_{z} } \left( \tau \hat{\sigma}_{x} \otimes \hat{s}_{y} + \hat{\sigma}_{y} \otimes \hat{s}_{x} \right) \exp^{ i \frac{\phi_{R}}{2} \hat{s}_{z} },
\end{equation}
where $\lambda_{R}$ describes the strength of the Rashba coupling, while the so-called Rashba angle, $\phi_{R}$, governs the spin-momentum locking of electron spins and therefore plays an essential role in the current-induced spin polarization effect (also known as Rashba-Edelstein effect).
In turn, the effect of sublattice-dependent proximity-exchange interaction is taken into account by the following term\cite{Zollner_PRB2016,Phong_Guinea_2018,Hogl_Gmitra_PRL2020}:
\begin{eqnarray}
\hat{H}_{EX} = \left(\lambda_{EX}^{A} \hat{\sigma}_{+}  - \lambda_{EX}^{B} \hat{\sigma}_{-}\right) \otimes \hat{s}_{z} ,
\end{eqnarray}
where $\lambda_{EX}^{A/B}$ are the sublattice dependent exchange coupling parameters. Finally, in the last term of Eq.~(\ref{eq:H_kp}), $E_0$ defines the position of the Dirac point with respect to the Fermi energy.

\begin{table}
\caption{The parameters defining the effective low-energy Hamiltonian of tGr/CrI$_3$, obtained from fitting to the  DFT data.}
\centering
\begin{tabular}{c c c c c c c c c c}
\hline 
\addlinespace
$a$ & $t$ & $E_0$ & $\Delta$ & $\lambda_{I}^{A}$ & $\lambda_{I}^{B}$ & $\lambda_{R}$ & $\phi_{R}$ & $\lambda_{EX}^{A}$ & $\lambda_{EX}^{B}$ \\
\,[\AA] & [eV] & [meV] & [meV] & [meV] & [meV] & [meV] & [rad] & [meV] & [meV] \\
\addlinespace
\hline
\addlinespace
2.62 & 2.02 & -5.7 & 5.7 & 0.8 & -0.99 & 1.65 & 3.75 & 16.01 & 20.79 \\
\addlinespace
\hline
\end{tabular}
\label{tab:tab2}
\end{table}

All parameters of Hamiltonian~(\ref{eq:H_kp}) can be obtained through a fitting procedure, in which the eigenstates of the $\mathbf{k}\cdot\mathbf{p}$ Hamiltonian are compared with the DFT-calculated electronic states of graphene around the $K$ and $K'$ points. In parallel with the band-structure fitting, the spin projections of electrons in the graphene electronic states are also matched to the DFT results. Table~\ref{tab:tab2} collects all the parameters obtained for tGr/CrI$_3$ by the fitting procedure. Figure~\ref{fig:Fig1sm} presents the electronic band structure and spin components in the vicinity of Dirac points, and shows a very good agreement between the effective model and DFT calculations.
The effective exchange coupling constants are about two orders of magnitude larger than in graphene/MnPSe$_3$~\cite{Hogl_Gmitra_PRL2020} and one order of magnitude larger than in graphene/Cr$_2$Ge$_2$Te$_6$~\cite{Zollner_PRL2022}. In turn, the intrinsic SOC and Rashba SOC parameters are comparable to those reported in semiconducting TMDCs~\cite{NaimerFabian2021,Zollner_PRB2023}. 

\section{Topological properties and selected transport characteristics}

\subsection{Berry curvature and Chern number}
Figures~\ref{fig:Fig2}(c),(e) reveal the inverted band structure of twisted graphene on CrI$_3$, indicating nontrivial topological properties of the heterostructure. To analyse this issue in more details, we determine  the Berry curvature, $\Omega_{\mathbf{k}}^{n}$, and Chern number, $n_{Ch}$. The Berry curvature corresponding to the $n$-th band can be obtained directly from the following  formula
\cite{Berry_RSLA1984,Thouless_PRL1982,Niu_Tholuess_PRB1985,Niu_PRB1996,XiaoRMP_2010,Sinova_RMP2010}:
\begin{equation}
\label{eq:BC}
\bOm_{\mathbf{k}n} = \nabla_{\mathbf{k}}\times\langle\Psi_{\mathbf{k}n}|i\nabla_{\mathbf{k}}|\Psi_{\mathbf{k}n}\rangle ,
\end{equation}
where $|\Psi_{\mathbf{k}n}\rangle$ is the basis function, i.e., the eigenfunction of the $\mathbf{k}\cdot\mathbf{p}$ Hamiltonian~(\ref{eq:H_kp}) or the cell-periodic Bloch state in the case of tight-binding or the first-principles calculations. Note that in the 2D system the Berry curvature has only the $z$ component, $\bOm_{\mathbf{k}n} = (0,0,\Omega_{\mathbf{k}n})$.
The Berry connection integrated over fully-occupied bands defines the Chern number \cite{Thouless_PRL1982,Niu_PRB1996}:
\begin{equation}
\label{eq:n_ch}
n_{Ch} = \frac{1}{2\pi} \sum_{\substack{n\\ \mathrm{occ.}}} \int d^{2}\mathbf{k}\,\Omega_{\mathbf{k}n}
\end{equation}
Note that in the case of a low-energy Hamiltonian, one needs to take into account the contribution from the two inequivalent K points, thus the Chern number calculated within the $\mathbf{k}\cdot\mathbf{p}$ model is:
\begin{equation}
\label{eq:n_ch_tau}
n_{Ch} = \sum_{\tau=\pm1} n_{Ch}^{\tau}
\end{equation}
and $n_{Ch}^{\tau}$ is defined by the Eq.~(\ref{eq:n_ch}) with Berry Curvature $\Omega_{\mathbf{k}n}^{\tau}$ calculated for the $\tau$-th valley.

\begin{figure}[t]
 \centering
\includegraphics[width=0.965\columnwidth]{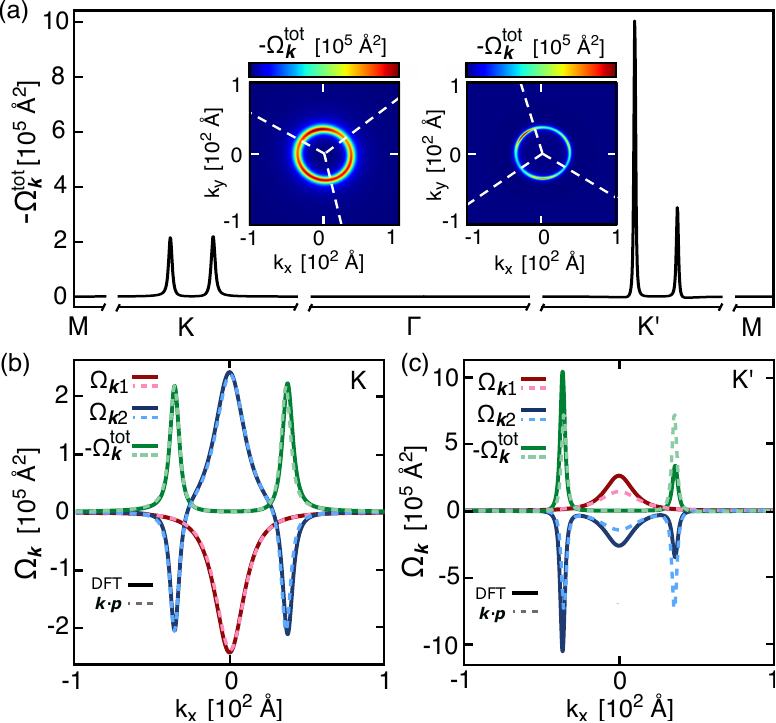}
 \caption{ Berry curvature of the t-Gr/CrI$3$ heterostructure for fully occupied valence bands. (a) DFT results obtained from maximally localized Wannier functions. (b),(c) Berry curvatures of individual valence bands of graphene and their sum obtained from the fitted low-energy model for the K (b) and K' (c) points.}
\label{fig:Fig5}
\end{figure}

Figure~\ref{fig:Fig5}(a) presents the total Berry curvature of the t-Gr/CrI$_3$ heterostructure, calculated from the DFT data for the fully occupied valence bands, i.e, for the Fermi level inside the energy gap of graphene. The Berry curvature was obtained from maximally localized Wannier functions using the \textsc{Wannier90} package~\cite{marzari1997maximally,mostofi2008wannier90} and the post-processing tools of \textsc{WannierTools}\cite{wu2018wanniertools}. The Berry curvatures for individual valence bands, calculated within the low-energy effective Hamiltonian, are shown in Fig.~\ref{fig:Fig5}(b,c) together with their sum (total Berry curvature). These results demonstrate not only the excellent agreement between the fitted $\mathbf{k}\cdot\mathbf{p}$ model and the DFT calculations, but also confirm that our low-energy Hamiltonian accurately captures the topological properties of the t-Gr/CrI$_3$ heterostructure.
\begin{figure}[t]
 \centering
\includegraphics[width=0.999\columnwidth]{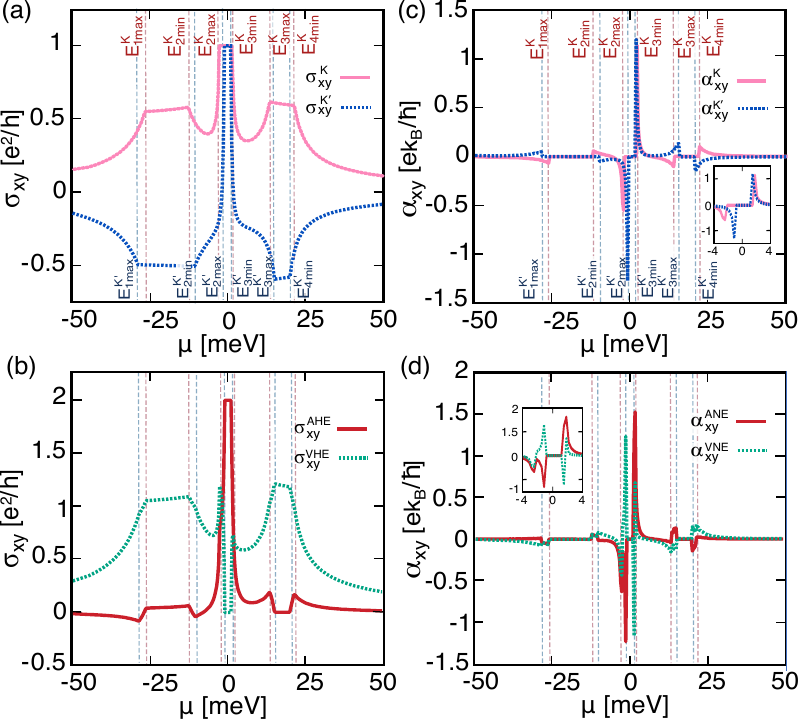}
 \caption{(a) The intrinsic contributions to the transverse conductivity, $\sigma_{xy}^{\rm K/K'}$, from the K and K' valleys, respectively.  (b) Anomalous Hall conductivity, $\sigma_{xy}^{\mathrm{AHE}}$, and the valley Hall conductivity, $\sigma_{xy}^{\mathrm{VHE}}$, plotted as a function of the chemical potential, calculated from the effective model. (c) Contributions to the transverse thermal conductivity, $\alpha_{xy}^{\rm K/K'}$, from the K and K' valleys, respectively. (d) Anomalous Nernst effect, $\alpha_{xy}^{\mathrm{ANE}}$, and valley Nernst effect, $\alpha_{xy}^{\mathrm{VNE}}$,  plotted as a function of the chemical potential. The $\mu = 0$ is set at the position of Dirac point, i.e.,  $\mu \rightarrow \mu - E_0$.}
\label{fig:Fig6}
\end{figure}

The total Berry curvature for fully occupied valence exhibits nonzero peaks around the K and K' points, where the graphene electronic states are located. Importantly, in consequence of time-reversal symmetry breaking, the peaks at the K and K' valleys have the same sign. The interplay between the proximity-induced exchange field and Rashba spin–orbit coupling opens an energy gap, which is further modulated by the staggered potential and intrinsic valley-Zeeman-type SOC. As a result, the insulating state in the t-Gr/CrI$_3$ heterostructure is topologically nontrivial. When the Fermi level is in the energy gap, the system is a Chern insulator, i.e., it hosts the quantum anomalous Hall (QAH) phase.
Indeed, the Chern numbers calculated for the K and K' valleys are identical, $n_{\mathrm{Ch}}^{\rm K} = n_{\mathrm{Ch}}^{{\rm K}'} = -1$, giving a total Chern number for the fully occupied valence bands of 
$n_{\mathrm{Ch}} = -2$. The same result is obtained directly from DFT calculations
for the Chern number of the fully occupied valence bands.
This is in agreement with earlier results on the topological properties of graphene on magnetic substrates~\cite{Qiao_PRB2010,Niu_PRB2012,Dyrdal_2DMat2017,Hogl_Gmitra_PRL2020,Inglot_PRB2021}.

\subsection{Anomalous and valley Hall effect}
The t-Gr/CrI$_3$ heterostructure exhibits Berry-phase-induced anomalous Hall effect (AHE). In the clean limit, the corresponding anomalous Hall conductivity can be found based on the Thouless-Kohmoto-Nightingale-Nijs (TKNN) theory~\cite{Thouless_PRL1982,Niu_Tholuess_PRB1985,Niu_PRB1996,XiaoRMP_2010}:
\begin{equation}
\sigma_{xy}^{\mathrm{AHE}} = -\frac{e^2}{h}\sum_{n}\int\frac{d^{2}\mathbf{k}}{2\pi} \Omega_{\mathbf{k}n}f(E_{\mathbf{k}n}),
\end{equation}
where the conductivity tensor $\sigma_{xy}$ is the ratio of charge current density flowing in the $x$-direction as a system response to the external electric field applied in the y-direction ($\sigma_{xy} \equiv j_{x}/E_y$) and  $f(E_{\mathbf{k}n})$ is the Fermi-Dirac distribution function for the $n$-th band.
Note that in the case of the $\mathbf{k}\cdot\mathbf{p}$ model, this expression should be treated as a sum of the contributions from both valleys (i.e., from K and K' points). Accordingly, $\sigma_{xy}^{\rm AHE}$ should read 
$\sigma_{xy}^{\rm AHE}\equiv\sigma_{xy}^{\rm K}+\sigma_{xy}^{{\rm K}'}$, as in this case $\Omega_{\mathbf{k}n} = \Omega_{\mathbf{k}n}^{\rm K} + \Omega_{\mathbf{k}n}^{{\rm K}'}$. 

The valley-contrasting behaviour of the band structure suggests the presence of a nonzero valley Hall effect (VHE)~\cite{Xiao_PRL2007,Ando_JPSocJap2015,Dyrdal_2DMat2017,Koppens_Sci2022}. This phenomenon follows directly from the valley-dependent anomalous velocity generated in an external electric field. Specifically, for a given band, the anomalous velocity components of electrons have opposite directions in the two valleys, as the corresponding Berry curvatures at the K and K' points have opposite signs. Consequently, electrons (or holes) from the two valleys are deflected toward opposite edges of the sample. The valley Hall conductivity is defined as
\begin{equation}
\sigma_{xy}^{\mathrm{VHE}} = \sigma_{xy}^{\rm K} - \sigma_{xy}^{{\rm K}'}.
\end{equation}
Importantly, in t-Gr/CrI$_3$ we have found that the bandgaps at the K and K' points differ by about 2~meV. In such a case, the bandgap can be closed selectively  (via external magnetic field or gate voltage) at one valley while remains open at the other one, realizing a valley-dependent bandgap closing. Otherwise, by tuning the chemical potential, one can obtain a current associated solely with the electronic states of a single valley.

Figure~\ref{fig:Fig6}(a) shows the Hall conductivities associated with the  K and K' points, whereas Fig.~\ref{fig:Fig6}(b) presents their sum, i.e. the anomalous Hall conductivity, $\sigma_{xy}^{{\mathrm{AHE}}}$, and their difference, that is the valley Hall conductivity $\sigma_{xy}^{\mathrm{VHE}}$ --  all presented as a function of the chemical potential. 
For chemical potentials within the gap (around $\mu = 0$), the anomalous Hall conductivity achieves the quantized and universal value $\sigma_{xy}^{\mathrm{AHE}} = 2e^{2}/h$. This quantization follows directly from the Chern number of the fully occupied valence bands of graphene, $n_{\mathrm{Ch}} = -2$. Outside the gap, the anomalous Hall conductivity rapidly decreases with increasing absolute value of the chemical potential. The kinks observed for both positive and negative chemical potentials correspond to the crossing of the valence and conduction band extrema at K and K' points, as indicated in Fig.~~\ref{fig:Fig6}. Furthermore, the asymmetry of the anomalous Hall conductivity with respect to the sign of the chemical potential reflects both particle–hole asymmetry and the inequivalence of the band extrema at the K and K' points. 

The VHE is absent, when the chemical potential is within the bandgap. However, when the chemical potential crosses the first conduction-band minimum ($E_{3\mathrm{min}}$) or the valence-band maximum ($E_{2\mathrm{max}}$), the valley Hall conductivity  exhibits a sharp peak, then decreases to a local minimum and increases again to its maximum value, when the chemical potential crosses the first conduction-band extremum ($E_{3\mathrm{max}}$) or the upper valence-band minimum ($E_{2\mathrm{min}}$), respectively. In the energy window between the local maximum of the first conduction band ($E_{3\mathrm{max}}$) and minimum of the second conduction band ($E_{4\mathrm{min}}$), as well as between the local minimum of the upper valence band ($E_{2\mathrm{min}}$) and the maximum of the lower valence band ($E_{1\mathrm{max}}$), we observe a plateau in the valance Hall conductivity

Here, we should note that  a modern theoretical framework for orbital effects has been proposed recently, suggesting that the valley Hall effect should, at some stage, be more rigorously described in terms of the orbital Hall effect~\cite{Johansson_JPCondMatrev2024}. Since the aim of this work is to highlight the potential of t-Gr/CrI$_3$ as an interesting platform for spintronics, we decided not to pursue this direction here. The orbital Hall effect in proximitized graphene has been discussed, for example, in~\cite{Vignale_PRB2021, Gobel_PRRes2023,Pezo_PRB2023,Sanchez_Canonico_PRRes2024}, and will be addressed in the context of graphene on magnetic substrates in future work.

\subsection{Anomalous and valley Nernst effect}
The Berry curvature is also responsible for transport phenomena driven by a temperature gradient in the system. One example is the anomalous Nernst effect (ANE), the thermal counterpart of the anomalous Hall effect. By analogy with the valley Hall effect, one can also define the valley Nernst effect (VNE). Both kinetic coefficients can be calculated from the following expressions:
\begin{equation}
\alpha_{xy}^{\rm ANE} = \alpha_{xy}^{\rm K}+\alpha_{xy}^{{\rm K}'},
\end{equation}
\begin{equation}
\alpha_{xy}^{\rm VNE} = \alpha_{xy}^{\rm K}-\alpha_{xy}^{{\rm K}'},
\end{equation}
where
\begin{equation}
\alpha_{xy}^{\tau} = \frac{e k_{B}}{\hbar} \sum_{n}\int\frac{d^2 \mathbf{k}}{(2\pi)^{2}} \Omega^{\tau}_{\mathbf{k}n}S^{\tau}_{\mathbf{k}n}.
\end{equation}
In the above equation $S_{\mathbf{k}n}$ is the entropy density of electrons, defined as:
\begin{eqnarray}
S_{\mathbf{k}n}^{\tau} = -f(E^{\tau}_{\mathbf{k}n})\ln(f(E^{\tau}_{\mathbf{k}n}))\hspace{2.0cm}\nonumber\\ - (1-f(E^{\tau}_{\mathbf{k}n}))\ln(1-f(E^{\tau}_{\mathbf{k}n})).
\end{eqnarray}

Figure~\ref{fig:Fig6}(c) presents the transverse thermal conductivity calculated for the K and K' points, respectively, whereas Fig~\ref{fig:Fig6}(d) presents the variation of ANE and VNE with chemical potential.  
Both characteristics exhibit sharp peaks at the chemical potential crossing the band extrema.
Accordingly, measurements of the anomalous or valley Hall effect, as well as of the anomalous or valley Nernst effect, allow for determination of both the bandgap width and the band splitting in proximitized graphene.

\begin{figure}[t]
 \centering
\includegraphics[width=0.99\columnwidth]{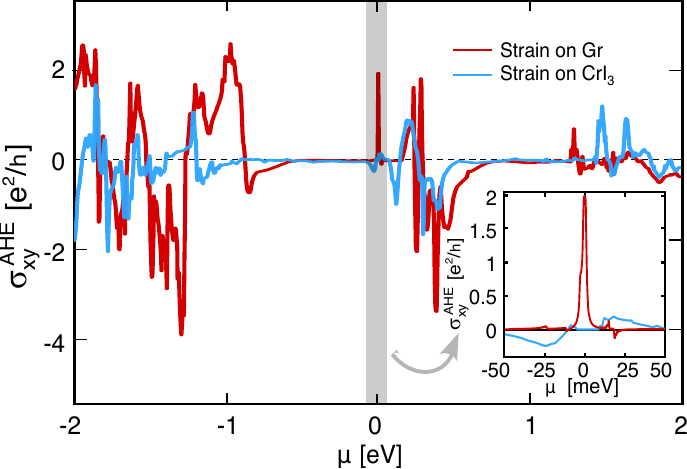}
 \caption{(a) Anomalous Hall conductivity calculated based on DFT data for t-Gr/CrI$_3$ when biaxial strain is applied only to graphene (red line) and when it is applied only to the CrI$_3$ monolayer. The $\mu = 0$ is set at the position of the Dirac point, i.e.,  $\mu \rightarrow \mu - E_0$.}
\label{fig:Fig7}
\end{figure}

\section{Effect of strain}

The electronic, topological, and magnetic properties of van der Waals materials are highly susceptible to mechanical strain~\cite{Kogl_npj2DMat2023,Liu_Strain_Valley_AdvFunctMat2023,Magnon_Straintronics_NanoLett2022,Yang_AdvOptMat2024}. It has been demonstrated that both the magnetic anisotropy energy and magnetic ground state of CrI$_3$ mono- and bi-layers can be modified by applying biaxial strain or gate voltage~\cite{Huang_NatNano2018,Leon_2DMat2020}. Accordingly, biaxial strain also modifies the net magnetization and the magnetic ground state of the t-Gr/CrI$_3$ heterostructure. These changes can be simply observed experimentally by the anomalous Hall conductivity measurement. In Fig.~\ref{fig:Fig7}, we present the anomalous Hall conductivity of the t-Gr/CrI$_3$ heterostructure, calculated from DFT data using the \textsc{Wannier90} package, for the two possible supercells that ensure commensurability between graphene and CrI$_3$ unit cells, as required for first-principles simulations. The red curves in Fig.~\ref{fig:Fig7}(a) and (b) correspond to the structure discussed throughout this work, in which biaxial strain is applied to the graphene monolayer while the CrI$_3$ monolayer remains unstrained. In this case, the system is in a ferromagnetic ground state and exhibits topologically nontrivial behavior, with a quantized anomalous Hall conductivity when the chemical potential lies within the energy gap. By contrast, the blue curves represent results for a supercell in which biaxial tensile strain of $\epsilon_{\mathrm{CrI_{3}}} = -5.90\%$ is applied to the CrI$_3$ layer, while the graphene remains unstrained. In this situation, we have found that the structure is in an antiferromagnetic ground state,  in agreement with previous studies reporting a magnetic phase transition under strain of $-5.5\%$ ~\cite{Zhang_PRB2018,Webster_PRB2018}. Importantly, the magnetic phase transition results in a strong modification of the proximity effects on the graphene electronic states. Consequently, t-Gr/CrI$_3$ becomes a trivial insulator with vanishing anomalous Hall conductivity when the chemical potential lies within the bandgap. 
We should note that the applied tensile strain is large and may cause the system to become unstable. This issue could potentially be mitigated by stabilizing the structure on a substrate or through encapsulation. However, such considerations (requiring DFT modelling of Gr/CrI$_3$ together with an additional substrate or encapsulating material) are beyond the scope of this work. Here, our objective was to emphasize the potential of magnetic van der Waals heterostructures for electronic and spintronic quantum devices, in which a strain-induced change in the quantum anomalous Hall conductivity from two conductance quanta to zero can encode logical “1” and “0,” respectively.

\section{Discussion and conclusions}

In this work, we provided a comprehensive analysis of the electronic, magnetic, and topological properties of graphene deposited on CrI$_3$. First, we have found the twist angle at which the graphene Dirac cones are positioned within the energy gap of CrI$_3$. Next, we have derived the low-energy effective Hamiltonian describing properties of the proximitized graphene,  which accurately captures the low-energy physics of the {t-Gr/CrI$_3$} heterostructure. We have also confirmed that the system is Chern insulator and reveals quantized anomalous Hall conductivity, when the chemical potential is placed inside the energy gap. We have analysed the intrinsic anomalous and valley Hall effect as well as anomalous and valley Nernst effects.
Characteristic kinks in the AHE/VHE (or sharp peaks in the ANE/VNE), that appear when the chemical potential is shifted from the energy gap into the first valence or conduction band, and when it crosses the edge of the second valence or conduction band, allow for determining the bandgap width and band splitting in proximitized graphene directly from transport measurements.

Another important feature of t-Gr/CrI$_3$ is the valley-contrasting behaviour of graphene Dirac cones. We have shown that the difference in the energy gap of the unbiased structure at the  K and K' valleys is approximately 2~meV. Accordingly, one can simply tune these gaps with an external gating or magnetic field. In such a case, the bandgap closes selectively at one valley, while remaining open at the other valley, enabling valley-dependent bandgap engineering and single-valley transport controlled by the chemical potential.
We have also highlighted the potential of magnetic van der Waals heterostructures for quantum electronic and spintronic devices, where strain can switch the quantum anomalous Hall conductivity between two conductance quanta and zero, encoding logical “1” and “0,” respectively.

We should also stress that in this article, we have presented transport coefficients that do not depend on the Rashba angle, $\phi_{R}$. The electronic states of twisted graphene on CrI$_3$ exhibit a large Rashba angle, $\phi_{R} = 3.75,\mathrm{rad}$ (approximately $215^{\circ}$), as shown in Fig.~\ref{fig:Fig2}(e) and (f). The Rashba angle substantially affects the nonequilibrium spin polarization, a phenomenon known also as the unconventional Rashba-Edelstein effect. 
A nonzero Rashba angle originates from the breaking of the in-plane mirror symmetry and results in spin–momentum locking, in which the electron spin expectation value acquires a finite radial component alongside the more commonly observed orthogonal component. Consequently, the nonequilibrium spin polarization has two in-plane components; one perpendicular and one parallel to the external electric field, as it was reported for graphene on transition metal dichalcogenides~\cite{Casanova_PRB22,Casanova_NatMat2024,NaimerFabian2021,Zollner_PRB2023,Wojciechowska_arxiv25}. In turn, it is also well known that in magnetic systems with strong spin-orbit coupling, the nonequilibrium spin polarization can possess all three components (depending on the magnetization orientation)~\cite{Dyrdal_PRB2015,Dyrdal_PRB2017,Krzyzewska_PhysE2022,Johansson_JPCondMatrev2024,Manchon_NatCommun2024}.
Accordingly, the nonequilibrium spin polarization emerging in t-Gr/CrI$_3$ will have various contributions that may, in slightly different ways, induce spin dynamics in the CrI$_3$ layer. A detailed theory of Rashba-Edelstein effect and spin-orbit torques in twisted graphene on ferromagnetic layers is discussed elsewhere~\cite{Wojciechowska_25tbp}.

\section*{acknowledgments}

M.J. and A.D. acknowledges the supported by the National Science Centre (NCN), Poland within the Norwegian Financial Mechanism under the Polish-Norwegian Research Project GRIEG ’2Dtronics’, project no. 2019/34/H/ST3/00515.

M.G. acknowledges funding by the EU NextGenerationEU through the Recovery and Resilience Plan for Slovakia under the project No. 09I05-03-V02-00071, and the Ministry of Education, Research, Development and Youth of the Slovak Republic, provided under Grant No. VEGA 1/0104/25, Slovak Research and Development Agency under Contract No. SK-SRB-23-0033, and the Slovak Academy of Sciences project IMPULZ IM-2021-42.

\bibliographystyle{apsrev4-2}
\bibliography{bib.bib}

\end{document}